\documentstyle[epsf]{aipproc}

\catcode`@=11
\long\def\@makefntext#1{\parindent 0pt\hsize\columnwidth\parskip0pt\relax
\footnotesize\baselineskip12pt\def\strut{\vrule width0pt height0pt depth1.75pt\relax}%
\mbox{$\m@th^{\@thefnmark}$\hspace*{3pt}}#1}

\def\title#1{\gdef\@title{{\par\vskip-10pt\Large\bf
\baselineskip20pt\centering\ignorespaces#1\vskip6pt}}%
\setcounter{part}{0}
\setcounter{table}{0}
\setcounter{figure}{0}
\setcounter{equation}{0}
\setcounter{section}{0}
\setcounter{subsection}{0}
\setcounter{subsubsection}{0}
\setcounter{paragraph}{0}
}

\def\author#1{\expandafter\def\expandafter\@authoraddress\expandafter
{\@authoraddress %
{\dimen0=-\prevdepth \advance\dimen0 by1.2\baselineskip
\nointerlineskip \centering
\vrule height\dimen0 width0pt\relax\ignorespaces\large\rm#1\par
}%
}%
}

\def\address#1{\expandafter\def\expandafter\@authoraddress\expandafter
{\@authoraddress{\nointerlineskip\vskip1pc
                 \footnotesize\it\centering\ignorespaces#1\par}}}

\catcode`@=12

\def\E{\,\rlap/\!E_T}
\def\srts{\sigma_{\!\!\!\sqrt s}^{\vphantom y}}

\newcommand{\lsim}{\mathrel{\raisebox{-.6ex}{$\stackrel{\textstyle<}{\sim}$}}}
\newcommand{\gsim}{\mathrel{\raisebox{-.6ex}{$\stackrel{\textstyle>}{\sim}$}}}

\begin{document}

\font\fortssbx=cmssbx10 scaled \magstep2
\hbox to \hsize{
\hbox{\fortssbx University of Wisconsin - Madison}
\hfill$\vcenter{\normalsize\hbox{\bf MADPH-98-1038}
                \hbox{February 1998}}$ }

\title{Supersymmetry vis-\`a-vis Muon Colliders\footnote{Invited talk presented at the {\it Workshop on Physics at the First Muon Collider and at the Front
End of a Muon Collider}, Fermilab, November 1997.}}
\author{V. Barger}
\address{Physics Department, University of Wisconsin, Madison, WI 53706, USA}

\maketitle

\pagestyle{plain}
\thispagestyle{empty}

\begin{abstract}
The potential of muon colliders to study a low-energy supersymmetry is addressed in the framework of the minimal supergravity model, whose predictions are first briefly surveyed. Foremost among the unique features of a muon collider is $s$-channel production of Higgs bosons, by which Higgs boson masses, widths, and couplings can be precisely measured to test the predictions of supersymmetry. Measurements of the threshold region cross sections of $W^+W^-$, $t\bar t$, $Zh$, chargino pairs, slepton and sneutrino pairs will precisely determine the corresponding masses and test supersymmetric radiative corrections. At the high-energy frontier a 3--4~TeV muon collider is ideally suited to study heavy scalar supersymmetric particles.
\end{abstract}

\section{Introduction}

There are indications that low energy supersymmetry (SUSY) is the right track for physics beyond the Standard Model (SM)\cite{reviews}. The measurements of gauge coupling strengths $\alpha_1, \alpha_2, \alpha_3$ are consistent with SUSY Grand Unification\cite{barbi} and global fits to precision electroweak measurements are consistent with SUSY expectations that there is a Higgs boson of mass less than 130~GeV\cite{langacker}. If nature is indeed supersymmetric, are there compelling arguments why muon colliders should be built? The answer is {\em YES}, and the reasons why are the subject of this report. The physics at muon colliders discussed herein is largely based on work in collaboration with M.~Berger, J.F.~Gunion, and T.~Han\cite{ourPR,moremumu}.

In the minimal supersymmetric model (MSSM) each standard model fermion (boson) has a boson (fermion) superpartner. Two Higgs doublets are required to give masses to the up-type and down-type fermions. SUSY breaking is introduced through all soft masses and couplings that do not introduce quadratic divergences; there are over 100 of these soft SUSY-breaking parameters. In SUSY-breaking models the number of independent soft parameters is greatly reduced. The breaking is transmitted from a hidden sector to the observable sector. There has been an explosive growth in models of SUSY breaking. These models fall into two classes, minimal SuperGravity (mSUGRA) and gauge-mediated symmetry breaking (GMSB). Thus, supersymmetry has many possible faces depending on:

\goodbreak
\begin{itemize}

\item breaking by general soft parameters or unification of soft parameters;

\item whether $R$-parity is conserved; with $R$-conservation the lightest supersymmetric particle (LSP) is stable;

\item whether the gravitino is heavy (in mSUGRA) or light (in GMSB);

\item the nature of the LSP: gaugino ($\tilde B$), higgsino ($\tilde H$), gravitino ($\tilde G$), or gluino ($\tilde g$);

\item the relative masses of the sparticles.

\end{itemize}

\noindent
Fortunately, there are some generic predictions that are not very model dependent. The most important is the guarantee of a light Higgs boson\cite{hhh}, which is accordingly the ``jewel in the crown" of supersymmetry. Also a light neutralino, chargino, sleptons and  sneutrinos are expected. In this report we concentrate on these lighter SUSY particles. 

\section{Rich SUSY Phenomenology}

In the mSUGRA model the neutralino $\tilde\chi_1^0$ is the LSP and it is a source of missing energy in SUSY events. The signatures of SUSY particle production in the mSUGRA model are leptons + jets + missing $E_T$ (denoted $\E$).

In the traditional GMSB models, the gravitino ($\tilde G$) is the LSP and it is a source of missing energy. The nature of the next-to-lightest supersymmetric particle (NLSP) in GMSB models determines the phenomenology. The NLSP options and their decays are $\chi_1^0\to\gamma\tilde G$, $\tilde\ell\to\ell\tilde G$ and $\tilde\tau_1\to\tau\tilde G$. Decays of the NLSP may occur within or outside the detector. The signatures of such events are $\gamma\gamma+\E$, $\ell\ell+\E$, \mbox{etc.\cite{kolda-susy}}

The mSUGRA model is the usual benchmark for SUSY phenomenology. The renormalization group equations relate masses and couplings at the scale of the Grand Unified Theory (GUT) to their electroweak scale values. The GUT scale parameter set in mSUGRA is $m_{1/2},\ m_0,\ \mu,\ A_0$ and $B_0$, where $m_{1/2}$ and $m_0$ are universal gaugino and scalar masses, $\mu$ is the Higgs mixing mass, $A_0$ is the trilinear coupling and  $B_0$ is the bilinear coupling. At the weak scale the phenomenology is determined by the parameters $m_{1/2},\ m_0,\ A_0,\ \rm sign(\mu)$ and $\tan\beta$, where $\tan\beta = v_u/v_d$ is the ratio of vacuum expectation values for the two Higgs doublets. A large top quark Yukawa coupling at the GUT scale is necessary to achieve electroweak symmetry breaking as a radiative effect\cite{inoue}. There are a number of predictions that follow from the large top Yukawa coupling:

\renewcommand{\labelenumi}{\theenumi)}
\begin{enumerate}

\item  The Higgs miracle happens at $M_Z$; electroweak symmetry breaking (EWSB) occurs radiatively.

\item  $m_b/m_\tau$ is correctly predicted\cite{aronson,bbo,bbo2} from $\lambda_b=\lambda_\tau$ unification\cite{chano} at the GUT scale.

\item $\lambda_t$ has an infrared fixed point\cite{bbo,bbo2,hill,bagger}, predicting (see Fig.~1)
$\tan\beta \simeq 1.8 \  (m_t = 200{\rm~GeV} \, \sin\beta)$ or $\tan\beta \simeq 56 \;.$

\begin{figure}[t]
\centering\leavevmode
\epsfxsize=3.8in\epsffile{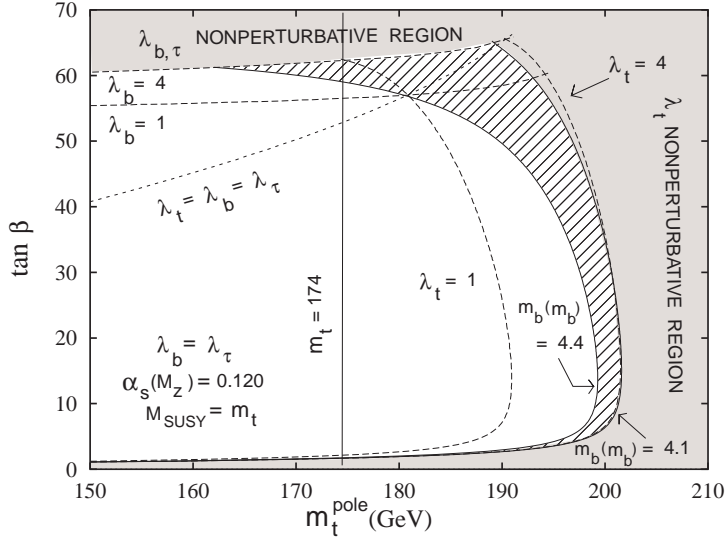}

\medskip
\caption[]{Contours of constant $m_b(m_b)$ in the $m_t(m_t), \tan\beta$ plane with contours of constant GUT scale Yukawa couplings. From Ref.~\cite{bbo}.}
\end{figure}

\item Gaugino masses at scale $M_Z$ are given by\cite{martin}
\begin{eqnarray}
&& M_1/\alpha_1 = M_2/\alpha_2 = M_3/\alpha_3 \;,\nonumber\\
&& M_1 = 0.44 m_{1/2}\,,\ M_2=0.88 m_{1/2}\,,\ M_3 = 3.2 m_{1/2} \;.\nonumber
\end{eqnarray}

\item $|\mu|$ is large compared to $M_1, M_2$ at scale $M_Z$; this follows from renormalization group evolution from the grand unification scale and minimization of the Higgs potential\cite{ellis}.

\item The chargino mass matrix has the approximate form\cite{ellis}
\[ {\cal M} \sim \left( \begin{array}{cc}
M_2 & 0 \\ 0 & -\mu
\end{array} \right) \mbox{ in the }
\left( \begin{array}{c} \tilde W^\pm\\ \tilde H^\pm \end{array} \right)\rm\ basis \,. \]
Thus $\tilde\chi_1^\pm\sim\tilde W^\pm$ and $\tilde\chi_2^\pm\sim\tilde H^\pm$. 

\item The neutralino mass matrix has the approximate form\cite{ellis} 
\[ {\cal M} = \left( \begin{array}{cccc}
M_1& 0 \\ 0& M_2 \\ & & 0& \mu\\ & & \mu& 0 \end{array} \right)
\mbox{ in the } \left( \begin{array}{c} 
\tilde B\\ \tilde W^3\\ \tilde H^0_d\\ \tilde H^0_u \end{array} \right) \rm\ basis \,. \]
Thus $\tilde\chi_1^0 \sim \tilde B^0$ and $\tilde\chi_2^0\sim \tilde W^3$.

\item The sparticle mass ratios are approximately in the proportions\cite{martin}
\[ \tilde\chi_1^0 : \tilde\chi_2^0 : \tilde\chi_1^\pm : g = 1:2:2:7 \;.\]
$\tilde\chi_1^0$ is the LSP.

\item The colored particles (squarks and the gluinos) are heavier; the scalar masses depend on $m_0$\cite{ellis}:	
\[ \tilde\chi_1^0, \tilde\chi_2^0, \tilde\chi_1^\pm, \tilde\ell, h \mbox{ are ``light''} \,,\]
\[ \tilde\chi_3^0, \tilde\chi_4^0, \tilde\chi_2^\pm, \tilde g, \tilde q \mbox{ are ``heavy''} \,.\]

\item The $\tilde\chi_1^0$ LSP is a natural candidate to explain the dark matter in the universe\cite{jungman}. The relic density $\Omega_{\tilde\chi_1^0}$ is inversely proportional to the thermally averaged $\tilde\chi_1^0\tilde\chi_1^0$ cross section, $\Omega_{\tilde\chi_1^0} h^2 \propto 1 / [\left<\sigma_{\rm ann} v\right>]$, where $h$ is the Hubble constant in units of 100~km/s/Mpc. The annihilation diagrams involve sfermion exchanges and $Z, h^0, H^0, A^0$ $s$-channel resonances. A cosmologically interesting LSP relic density, $0.1\lsim \Omega_{\tilde\chi_1^0} h^2 \lsim 0.5$, singles out the following region of mSUGRA \mbox{parameters\cite{ourDM,howie}:}
\[\begin{array}{lcl}
 m_0 \lsim 200{\rm~GeV},& 80\lsim m_{1/2} \lsim 450{\rm~GeV} & \rm for\ \tan\beta\sim 1.8 \;,\\
 m_0 \gsim 300{\rm~GeV},& 500\lsim m_{1/2} \lsim 800{\rm~GeV} & \rm for\ \tan\beta\sim50 \,.
\end{array}\]
The sparticle mass spectra are correspondingly constrained; see Fig.~2.

\begin{figure}[h]
\centering\leavevmode
\epsfxsize=3.8in\epsffile{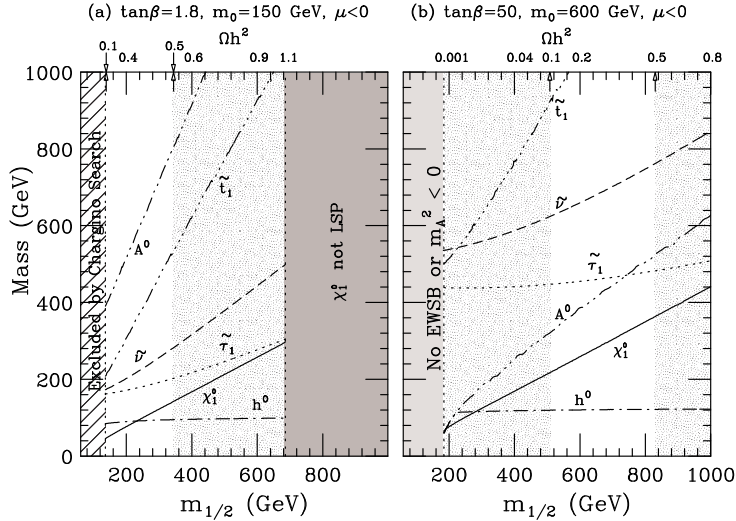}

\caption[]{ The neutralino relic density 
and SUSY Higgs mass spectrum versus $m_{1/2}$ for $\mu > 0$ 
with (a)~$\tan\beta = 1.8$, $m_0 = 150$ GeV and  
(b)~$\tan\beta = 50$, $m_0 = 600$ GeV.
$\tilde{\nu}$ is the lightest scalar neutrino.
The shaded regions denote the parts of the parameter space 
(i)~producing $\Omega_{\chi^0_1} h^2 < 0.1$ or $\Omega_{\chi^0_1} h^2 > 0.5$, 
(ii)~excluded by theoretical requirements, 
or (iii)~excluded by the chargino search at \mbox{LEP-2}. From Ref.~\cite{ourDM}.}
\end{figure}

\item The mass of the lightest MSSM Higgs boson ($h^0$) is bounded from above. At tree level $m_h\le M_Z |\cos2\beta|$\cite{hhg}. Radiative corrections from top and stop loops increase the bound to $m_h\lsim130$~GeV\cite{hhh}. In models with extra Higgs doublets and singlets that remain perturbative to the Planck scale, at least one Higgs boson has mass $m_h\lsim 150$~GeV\cite{kolda}. Thus the lightest Higgs boson is a secure target of a low-energy supersymmetry. 

\item The other Higgs bosons have masses\cite{ourDM,hightanb}
\[ \begin{array}{lll}
m_A \approx m_H \gg M_Z& \rm if& \tan\beta\sim1.8\;,\\
m_A \sim{\cal O}(M_Z)& \rm if& \tan\beta\gsim m_t/m_b
\mbox{ or }m_0\sim50\rm~GeV\;.
\end{array} \]

\end{enumerate}

\section{Higgs Physics at a Muon Collider}

The production of Higgs bosons in the $s$-channel with interesting rates is an unique feature of a muon collider\cite{ourPR,bbgh}. The resonance cross section is
\begin{equation}
\sigma_h(\sqrt s) = {4\pi \Gamma(h\to\mu\bar\mu) \, \Gamma(h\to X)\over
\left(\hat s - m_h^2\right)^2 + m_h^2 \left(\Gamma_{\rm tot}^h \right)^2}
\end{equation}
Gaussian beams with root mean square resolution  down to $R=0.003\%$ are realizable. The corresponding root mean square spread $\srts$ in c.m.\ energy is
\begin{equation}
\srts = (2{\rm~MeV}) \left( R\over 0.003\%\right) \left(\sqrt s\over 100\rm~GeV\right) \,.
\end{equation}
The effective $s$-channel Higgs cross section convolved with a Gaussian spread
\begin{equation}
\bar\sigma_h(\sqrt s) = {1\over \sqrt{2\pi}\,\srts} \; \int \sigma_h (\sqrt{\hat s}) \; \exp\left[ -\left( \sqrt{\hat s} - \sqrt s\right)^2 \over 2\sigma_{\sqrt s}^2 \right] d \sqrt{\hat s} 
\end{equation}
is illustrated in Fig.~3 for $m_h = 110$~GeV, $\Gamma_h = 2.5$~MeV, and resolutions $R=0.01\%$, 0.06\% and 0.1\%\cite{ourPR,bbgh}. A resolution $\srts \sim \Gamma_h$ is needed to be sensitive to the Higgs width. The light Higgs width is predicted to be\cite{bbgh}
\begin{equation}
\begin{array}{lll}
\Gamma \approx 2\mbox{ to 3 MeV}& \rm if& \tan\beta\sim1.8\\
\Gamma \approx 2\mbox{ to 800 MeV}& \rm if& \tan\beta\sim20
\end{array}
\end{equation}
for $80{\rm~GeV}\lsim m_h\lsim120$~GeV.

At $\sqrt s = m_h$, the effective $s$-channel Higgs cross section is\cite{ourPR}
\begin{equation}
\bar\sigma_h \simeq {4\pi\over m_h^2} \; {{\rm BF}(h\to\mu\bar\mu) \,
{\rm BF}(h\to X) \over \left[ 1 + {8\over\pi} \left(\srts\over\Gamma_{\rm tot}^h\right)^2 \right]^{1/2}} \,.
\end{equation}
Note that $\bar\sigma_h\propto 1/\srts$ for $\srts>\Gamma_{\rm tot}^h$. At $\sqrt s = m_h \approx 110$~GeV, the $b\bar b$ rates are\cite{ourPR,bbgh}
\begin{eqnarray}
\rm signal &\approx& 10^4\rm\ events/fb\\
\rm background &\approx& 10^4\rm\ events/fb
\end{eqnarray}
assuming a $b$-tagging efficiency $\epsilon \sim 0.5$. The effective on-resonance cross sections for other $m_h$ values and other channels ($ZZ^*, WW^*$) are shown in Fig.~4 for the SM Higgs. The rates for the MSSM Higgs are nearly the same as the SM rates in the decoupling regime\cite{decoupl}, which is relevant at $\tan\beta\sim1.8$ in mSUGRA.

\begin{figure}[h]
\centering\leavevmode
\epsfxsize=3in\epsffile{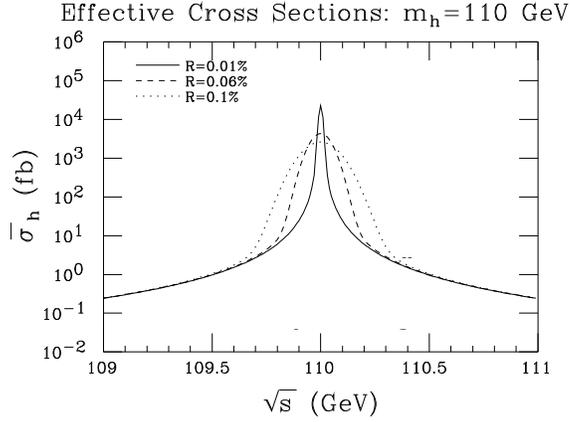}

\caption[]{Effective $s$-channel higgs cross section $\bar\sigma_h$ obtained by convoluting the Breit-Wigner resonance formula with a Gaussian distribution for resolution $R$. From Ref.~\cite{ourPR}.}
\end{figure}

\begin{figure}[h]
\centering\leavevmode
\epsfxsize=4in\epsffile{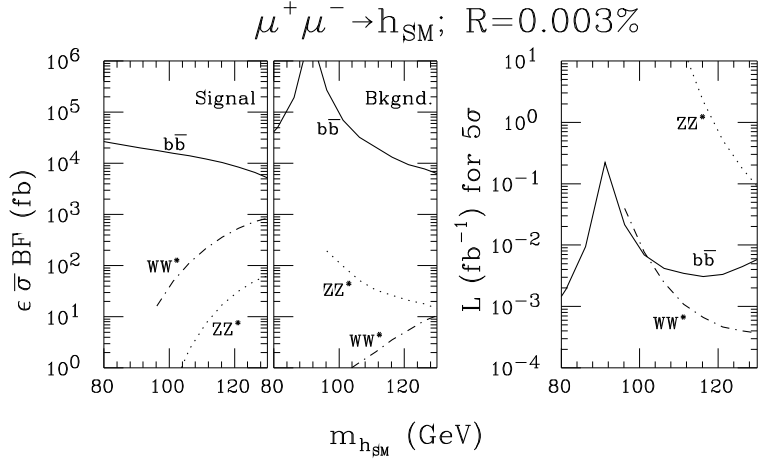}

\caption[]{The SM Higgs cross sections and backgrounds in $b\bar b,\ WW^*$ and $ZZ^*$. Also shown is the luminosity needed for a 5~standard deviation detection in $b\bar b$. From Ref.~\cite{ourPR}.}
\end{figure}

The important factors that make $s$-channel Higgs physics studies possible at a muon collider are energy resolutions $\srts$ of order a few MeV, little bremsstrahlung and no beamstrahlung smearing, and precise tuning of the beam energy to an accuracy $\Delta E\sim10^{-6}E$ through continuous spin-rotation measurements\cite{raja}. As a case study we discuss $m_h \approx 110$~GeV. Prior Higgs discovery is assumed at the Tevatron (in $Wh, t\bar th$ with $h\to b\bar b$), at the LHC (in $gg\to h$ with $h\to \gamma\gamma, 4\ell$ with a mass measurement of $\Delta m_h \sim 100$~MeV for an integrated luminosity of $L=300\rm~fb^{-1}$) or possibly at a NLC (in $Z^*\to Zh, h\to b\bar b$ giving $\Delta m_h \sim 50$~MeV for $L=200\rm~fb^{-1}$). A muon collider ring design would be optimized to run at energy $\sqrt s= m_h$. For an initial Higgs mass uncertainty of $\Delta m_h\sim 100$~MeV, the maximum number of scan points required to locate the $s$-channel resonance peak at the muon collider is
\begin{equation}
n = 2\Delta m/\srts \approx 100
\end{equation}
for a resolution $\srts \approx 2$~MeV. The necessary luminosity per scan point ($L_{\rm s.p.}$) to observe or eliminate the $h$-resonance at a significance level $S/\sqrt B = 3$ is $L_{\rm s.p.} \sim 1.5\times10^{-3}\,\rm fb^{-1}$. (The scan luminosity requirements increase for $m_h$ closer to $M_Z$; at $m_h\sim M_Z$ the $L_{\rm s.p.}$ needed is a factor of 50 higher.) The total luminosity then needed to tune to a Higgs boson with $m_h = 110$~GeV is $L_{\rm tot} = 0.15\rm~fb^{-1}$. If the machine delivers  $5\times10^{30}\rm\, cm^{-2}\, s^{-1}$ (0.05 fb$^{-1}$/year), the luminosity criteria specified for this workshop, then 3 years running would be needed. However, luminosities of order $1.5\times10^{31}\rm\,cm^{-2}\,s^{-1}$ are currently believed to be realizable\cite{palmer}, in which case only one year of running would suffice to complete the scan and measure the Higgs mass to an accuracy  $\Delta m \sim 1$~MeV. Figure~5 illustrates a simulation of such a scan. 

\begin{figure}[h]
\centering\leavevmode
\epsfxsize=3in\epsffile{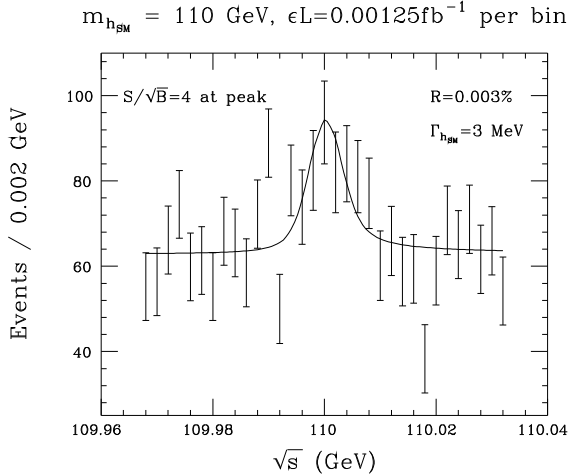}

\caption[]{Number of events and statistical errors in the $b\bar b$ final states as a function of $\sqrt s$ in the vicinity of $m_{h_{\rm SM}}=100$~GeV, assuming $R=0.03\%$. From Ref.~\cite{ourPR}.}
\end{figure}

Once the $h$-mass is determined to $\sim1$~MeV, a 3-point fine scan\cite{ourPR} can be made across the peak with higher luminosity, distributed with $L_1$ at the observed peak position in $\sqrt s$ and $2.5L_1$ at the wings ($\sqrt s = {\rm peak} \pm 2\srts$). Then with $L_{\rm tot}= 0.4\rm~fb^{-1}$ the following accuracies would be achievable: 16\% for $\Gamma_{\rm tot}^h$, 1\% for $\sigma\rm BF(b \bar b)$ and 5\% for $\sigma\rm BF(WW^*)$. The ratio $r = {\rm BF}(WW^*)/ {\rm BF} (b\bar b)$ is sensitive to $m_A$ for $m_A$ values below 500~GeV. For example, $r_{\rm MSSM}/r_{\rm SM} = 0.3, 0.5, 0.8$ for $m_A = 200, 250, 400$~GeV\cite{ourPR}. Thus, it may be possible to infer $m_A$ from $s$-channel measurements of $h$.

The study of the other neutral MSSM Higgs bosons at a muon collider via the $s$-channel is also of major interest. Finding the $H^0$ and $A^0$ may not be easy at other colliders. At the LHC the region $m_A>200$~GeV is deemed to be inaccessible for $3\lsim\tan\beta\lsim5$--10. At an NLC the $e^+e^-\to H^0 A^0$ production process may be kinematically inaccessible if $H^0$ and $A^0$ are heavy. At a $\gamma\gamma$ collider, very high luminosity (${\sim}200\rm\ fb^{-1}$) would be needed for $\gamma\gamma\to H^0, A^0$ studies. At a muon collider the resolution requirements for $s$-channel $H^0$ and $A^0$ studies are not as demanding as for the $h$, because the $H^0, A^0$ widths are broader; typically $\Gamma\sim30$~MeV for $m_A<2m_t$ and $\Gamma\sim3$~GeV for $m_A>2m_t$. Consequently $R\sim0.1\%$ ($\srts \sim 70$~MeV) is adequate for a scan. A luminosity per scan point $L_{\rm s.p.}\sim 0.1\rm~fb^{-1}$ probes the parameter space with $\tan\beta>2$.
 The $\sqrt s$-range over which the scan should be made depends on other information available to indicate the $A^0$ and $H^0$ mass ranges of interest.

In mSUGRA with large $m_A$, $m_{A^0}\approx m_{H^0}\approx m_{H^\pm}$ and the degeneracy in these masses is very close for large $\tan\beta$. In such a circumstance only an $s$-channel scan with good resolution may allow separation of the $A^0$ and $H^0$ states; see Fig.~6.

\begin{figure}[h]
\centering\leavevmode
\epsfxsize=3in\epsffile{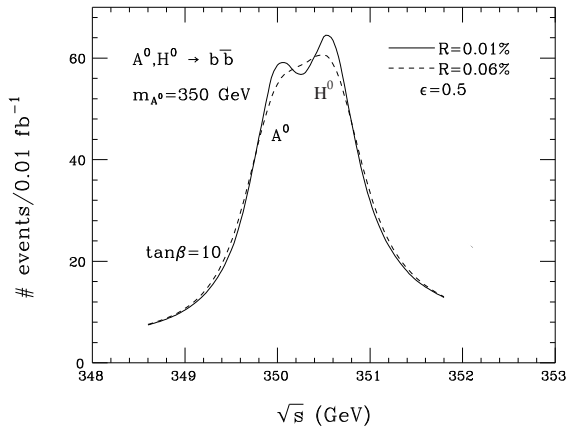}

\caption[]{Separation of $A^0$ and $H^0$ signals for $\tan\beta=10$. From Ref.~\cite{ourPR}. }
\end{figure}

\section{Threshold Measurements at a Muon Collider}

With 10~fb$^{-1}$ integrated luminosity devoted to a measurement of a threshold cross-section, the following precisions on particle masses may be achievable\cite{bbgh2,bbh-new}:
\begin{equation}
\begin{array}{ll}
\mu^+\mu^-\to W^+W^-& \Delta M_W = 20\rm\ MeV\,,\\
\mu^+\mu^-\to t\bar t & \Delta m_t = 0.2\rm\ GeV\,,\\
\mu^+\mu^-\to Zh& \Delta m_h = 140{\rm\ MeV\ \ (if\ } m_h = 100\rm\ GeV)\,.
\end{array}
\end{equation}
Precision $M_W$ and $m_t$ measurements allow important tests of electroweak radiative corrections through the relation
\begin{equation}
M_W = M_Z \left[ 1 - {\pi\alpha\over \sqrt 2 \, G_\mu \, M_W^2 (1-\delta r)} \right]^{1/2} \,,
\end{equation}
where $\delta r$ represents loop corrections. In the SM, $\delta r$ depends on $m_t^2$ and $\log m_h$. The optimal precision for tests of this relation is $\Delta M_W \approx {1\over 140}\Delta m_t$, so the uncertainty on $M_W$ is the most critical. With $\Delta M_W=20$~MeV the SM Higgs mass could be inferred to an accuracy 
\begin{equation}
\Delta m_{h_{\rm SM}} = \pm 30{\rm\ GeV} \left(m_h\over 100\rm\ GeV\right)\,.
\end{equation}
Alternatively, once $m_h$ is known from direct measurements, SUSY loop contributions can be tested.

One of the important physics opportunities for the First Muon Collider is the production of the lighter chargino, $\tilde\chi_1^+$. Fine tuning arguments in mSUGRA suggest that it should be lighter than 200~GeV\cite{chen}. A search at the upgraded Tevatron for the process $q\bar q\to\tilde\chi_1^+\tilde\chi_2^0$ with $\tilde\chi_1^+\to \tilde\chi_1^0\ell^+\nu$ and $\tilde\chi_2^0\to\tilde\chi_1^0\ell^+\ell^-$ decays can reach masses $m_{\tilde\chi_1^+}\simeq m_{\tilde\chi_2^0}\sim 170$~GeV with 2~fb$^{-1}$ luminosity and $\sim230$~GeV with 10~fb$^{-1}$\cite{teva2000}. The mass difference $M(\tilde\chi_2^0) - M(\tilde\chi_1^0)$ can be determined from the $\ell^+\ell^-$ mass distribution.

The two contributing diagrams in the chargino pair production process are shown in Fig.~7; the two amplitudes interfere destructively\cite{feng}. The $\tilde\chi_1^+$  and $\tilde\nu_\mu$ masses can be inferred from the shape of the cross section in the threshold region\cite{bbh-new}. The chargino decay is $\tilde\chi_1^+\to f\bar f' \tilde\chi_1^0$. Selective cuts suppress the background from $W^+W^-$ production and leave $\sim5\%$ signal efficiency for 4\,jets${}+\E$ events. Measurements at two energies in the threshold region with total luminosity $L=50\rm~fb^{-1}$ and resolution $R=0.1\%$ can give the accuracies listed in Table~1 on the chargino mass for the specified values of $m_{\tilde\chi_1^+}$ and $m_{\tilde\nu_\mu}$. 

\begin{figure}[h]
\centering\leavevmode
\epsfxsize=3.5in\epsffile{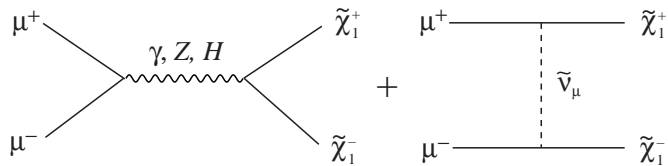}

\caption{Diagrams for production of the lighter chargino.}
\end{figure}

\begin{table}[t]
\caption[]{Achievable uncertainties with 50~fb$^{-1}$ luminosity on the mass of the lighter chargino for representative $m_{\tilde\chi_1^+}$ and $m_{\tilde\nu_\mu}$ masses. From Ref.~\cite{bbh-new}. }
\begin{tabular}{c}
\hfill\epsfxsize=3in\epsffile{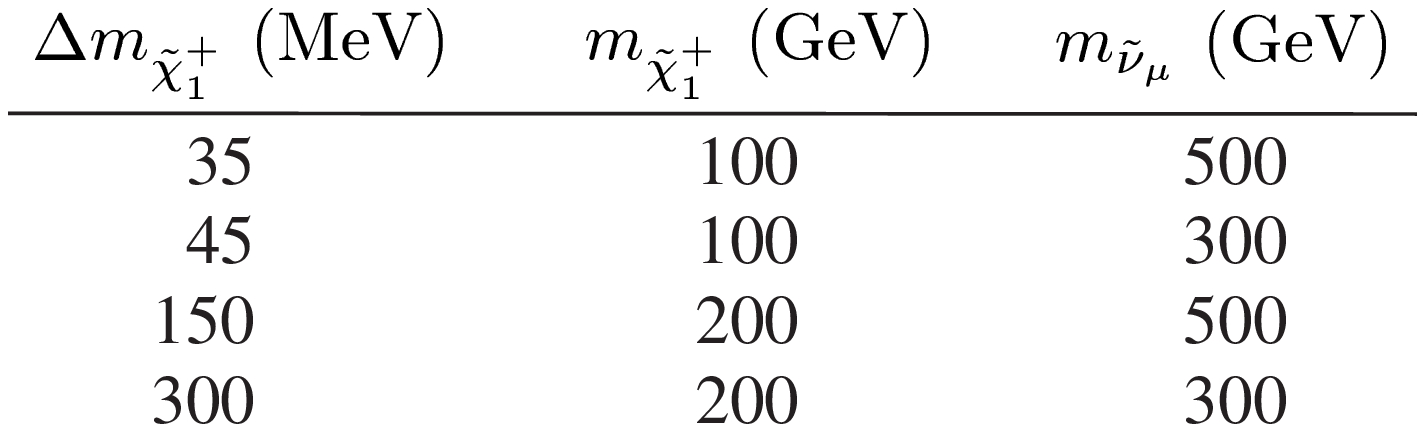}\hfil
\end{tabular}
\end{table}

\section{Supersymmetric Radiative Corrections}

In unbroken supersymmetry, the SUSY gaugino couplings $h_i$ to $\tilde ff$ are equal to the SM gauge couplings $g_i$. In broken SUSY a difference in $h_i$ and $g_i$ couplings is induced at the loop level due to different mass scales for squarks and sleptons\cite{chank,nojiri,cheng,diaz}. The differences in the U(1) and SU(2) couplings are\cite{cheng}
\begin{eqnarray}
{h_1-g_1\over g_1} &\simeq& 1.8\% \log_{10} \left(M_{\tilde Q}\over m_{\tilde \ell} \right) \,,\\
{h_2-g_2\over g_2} &\simeq& 0.7\% \log_{10} \left(M_{\tilde Q}\over m_{\tilde \ell}\right) \,.
\end{eqnarray}
One-loop amplitudes for SUSY processes are obtained from the tree-level amplitudes by substitution of the modified couplings. The cross-sections of SUSY processes with $t$-channel exchanges can be enhanced up to $9\%\log_{10} \left( M_{\tilde Q}/m_{\tilde\ell}\right)$\cite{nojiri}. Consequently, precision cross-section measurements can be sensitive to squarks of mass $M_{\tilde Q}>1$~TeV. If the first two generations have masses in the 1 to 40~TeV range allowed by naturalness, then precision measurements could provide a way to infer squark masses beyond the kinematic reach of colliders.

Some $t$-channel exchange processes of interest in this regard at muon colliders are shown in Fig.~8. The technique relies on knowledge of the exchanged particle mass, which must be determined from its production processes. The muon collider advantage in the study of supersymmetric radiative corrections is the accuracy with which mass measurements can be made near thresholds. 

\begin{figure}[t]
\centering\leavevmode
\epsfxsize=4in\epsffile{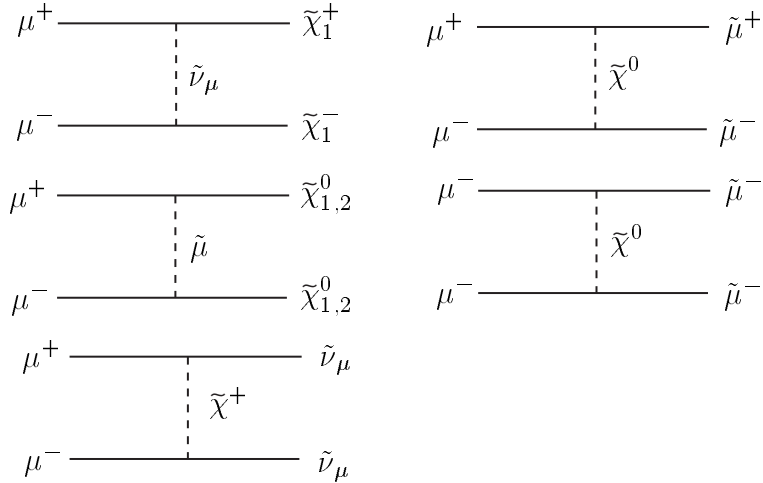}

\caption{$t$-channel exchange diagrams for processes that can be enhanced by SUSY radiative corrections.}
\end{figure}

\section{Heavy SUSY Particles}

Of the many scalar particles of supersymmetry (sleptons, squarks, Higgs) some may have masses of TeV scale. Study of heavy SUSY particles at the LHC will be difficult because of low event rates and high SM backgrounds. At a lepton collider pair production of scalars is $p$-wave suppressed. Consequently, collider energies well above threshold are necessary to have sufficient production rates; see Fig.~9. A~3~to 4~TeV muon collider offers the promise of high luminosity ($\sim1000\rm~fb^{-1}/year$) that would allow sufficient event rates to reconstruct heavy sparticles from their complex cascade decay chains. 

\begin{figure}
\centering\leavevmode
\epsfxsize=5.5in\epsffile{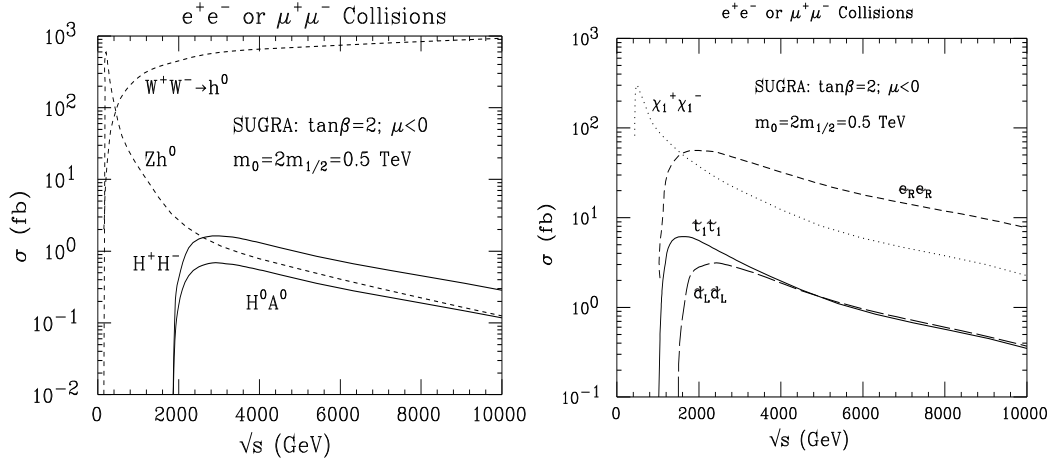}

\caption[]{Cross sections for pair production of Higgs bosons and scalar particles at a high energy muon collider. From Ref.~\cite{sanfran95}.}
\end{figure}

\section{Conclusions}

Muon colliders offer unique probes of supersymmetry. The $s$-channel production of Higgs bosons will precisely determine the Higgs mass (to a fraction of an MeV), directly measure the Higgs width, measure the branching fraction ratio BF$(h\to WW^*)/$BF$(h\to b\bar b)$ from which $m_A$ can be inferred if $m_A<500$~GeV, and allow discovery and study of the $A^0$ and $H^0$ Higgs bosons. High precision threshold cross-section measurements are possible at a muon collider because of the sharp beam resolution, the suppressed bremsstrahlung and the precise tuning of the beam energy through spin-rotation measurements. Interesting possibilities for precise mass measurements at the First Muon Collider include:
\begin{equation}
\begin{array}{ll}
W^+W^-, t\bar t & (M_W, m_t)\\
Zh & (m_h)\\
\tilde\chi^+\tilde\chi^-, \tilde\chi_{1,2}^0 \, \tilde\chi_{1,2}^0 &
(m_{\tilde\chi^+}, m_{\tilde\chi_1^0}, m_{\tilde\chi_2^0})\\
\tilde\nu_\mu\tilde\nu_\mu & (m_{\tilde\nu_\mu})\\
\tilde\ell^+\tilde\ell^- & (m_{\tilde\ell})
\end{array}
\end{equation} 
Precision cross-section measurements may allow tests of supersymmetric radiative corrections that may allow us to infer the existence of squarks with mass above 1~TeV. Finally, the next generation muon collider with c.m.\ energy of 3 to 4~TeV would provide access to heavy SUSY particles.

The bottom line is that muon colliders are a robust option for discovering  the nature of supersymmetry.

\section*{Acknowledgments}
I would like to thank M.S.~Berger, J.F.~Gunion, T.~Han and C.~Kao for helpful advice in the preparation of this report. This research was supported in part by the U.S.~Department of Energy under Grant No.~DE-FG02-95ER40896 and in part by the University of Wisconsin Research Committee with funds granted by the Wisconsin Alumni Research Foundation.

\end{document}